# Cold Atmospheric Plasma discharged in Water and its Potential Use in Cancer Therapy

Zhitong Chen, Xiaoqian Cheng, Li Lin, Michael Keidar[*]

Department of Mechanical and Aerospace Engineering, The George Washington University,

Washington, DC 20052, USA

**Abstract**

Cold atmospheric plasma (CAP) has been emerged as a novel technology for cancer treatment. CAP can directly treat cells and tissue but such direct application is limited to skin or can be invoked as a supplement during open surgery. In this study we report indirect plasma treatment using CAP discharged in DI water using three gases as carriers (argon, helium and nitrogen). Plasma stimulated water was applied to human breast cancer cell line (MDA-MB-231). MTT assay tests showed that using argon plasma had the strongest effect on inducing apoptosis in cultured human breast cancer cells. This result is attributed to the elevated production of the reactive oxygen species and reactive nitrogen species in water in the case of argon plasma.

---

[*] Corresponding Author:
E–mail address: zhitongchen@gwu.edu, keidar@gwu.edu

# 1. Introduction

The most important aspect of successful cancer therapy is the selective eradication of cancer cell without influencing the healthy tissue. In the course of cell's normal activity, various reactive oxygen species (ROS) and reactive nitrogen species (RNS) are produced[1,2]. It is well known that ROS and RNS are able to induce cell proliferation as well as cell death, while extreme amounts of reactive oxygen and nitrogen species (RONS) can induce apoptosis and damage of proteins, lipids, DNA[3,4]. Recently, cold atmospheric plasma (CAP) has been introduced as a new tool with a potential on the current paradigm of cancer treatment[5,6]. Moreover, CAP has shown significantly potential for various biomedical applications such as sterilization of infected tissues[7,8], inactivation of microorganisms[9], wound healing[10,11], skin regeneration[12], blood coagulation[13], tooth bleaching[14], and cancer therapy[15-18]. CAP has been well known for the generation of RONS that affect various cellular functions. It should be pointed out that CAP produces similar level of unique chemical compositions and reactions to endogenous RONS cell chemistry. In addition to various RONS, CAP produces UV, charged particles (electrons, ions), and electric field[19-21]. CAP had significant success both in vitro and in vivo studies on cancer therapy applications, including breast cancer[4], lung carcinoma[5,22,23], hepatocellular carcinoma [24], neuroblastoma[25], skin carcinoma[26], melanoma[27], colon carcinoma[28], pancreatic carcinoma[29], bladder carcinoma[5], and cervical carcinoma[30,31], et al.

In most reported studies CAP was used directly to irradiate cancer cells or tissues. However, CAP jet direct irradiation can only cause cell death in the upper 3-5 cell layers[32]. Meanwhile, such direct application is limited to skin or can be invoked as a supplement during open surgery. In this study, we report that discharge submerged in water (plasma solution) has more advantages in comparison with direct treatment. Blood constitutes 55% of blood fluid, which is mostly water (92% by volume). It is very difficult to directly treat blood by CAP due to its coagulation and higher viscosity coefficient than that of the water. Thus, we can use strategy of plasma solution and injecting it into the blood around tumor. On the other hand, we envision that plasma solutions might be used as a basis for a drug against digestive system cancer. To this end, we treated DI water by CAP and studied its effects on cancer cells. In this study, a new CAP device had been designed to generate RONS in deionized water (DI). Such CAP stimulated DI water can be applied to cells and tissues when direct access by plasma is not available. This study describes the plasma device, its characterization and response of cancer cell on plasma-stimulated DI water application.

## 2. Experimental section

*2.1 CAP device immerged in DI water*

The CAP device immerged in DI water is shown in Fig. 1a, b. It consists of 2 electrodes assembly with a central powered electrode (1 mm in diameter) and a grounded outer electrode wrapped around the outside of quartz tube (4.5 mm in diameter). The electrodes were connected to a secondary output of high voltage transformer (voltage in the range of 2-5 kV, frequency of about 30 kHz). Industrial grade Nitrogen, Argon, and Helium were used as carrier gases. The flow rate was maintained at about 0.3 L/min. CAP treatment outside of DI water was also used for comparison. Images of plasma generated in DI water using $N_2$, He and Ar as carrier gases are shown in Fig.1 (d). One can see that the plasma generated in DI water using $N_2$ is very weak, while Ar generated plasma has the strongest intensity.

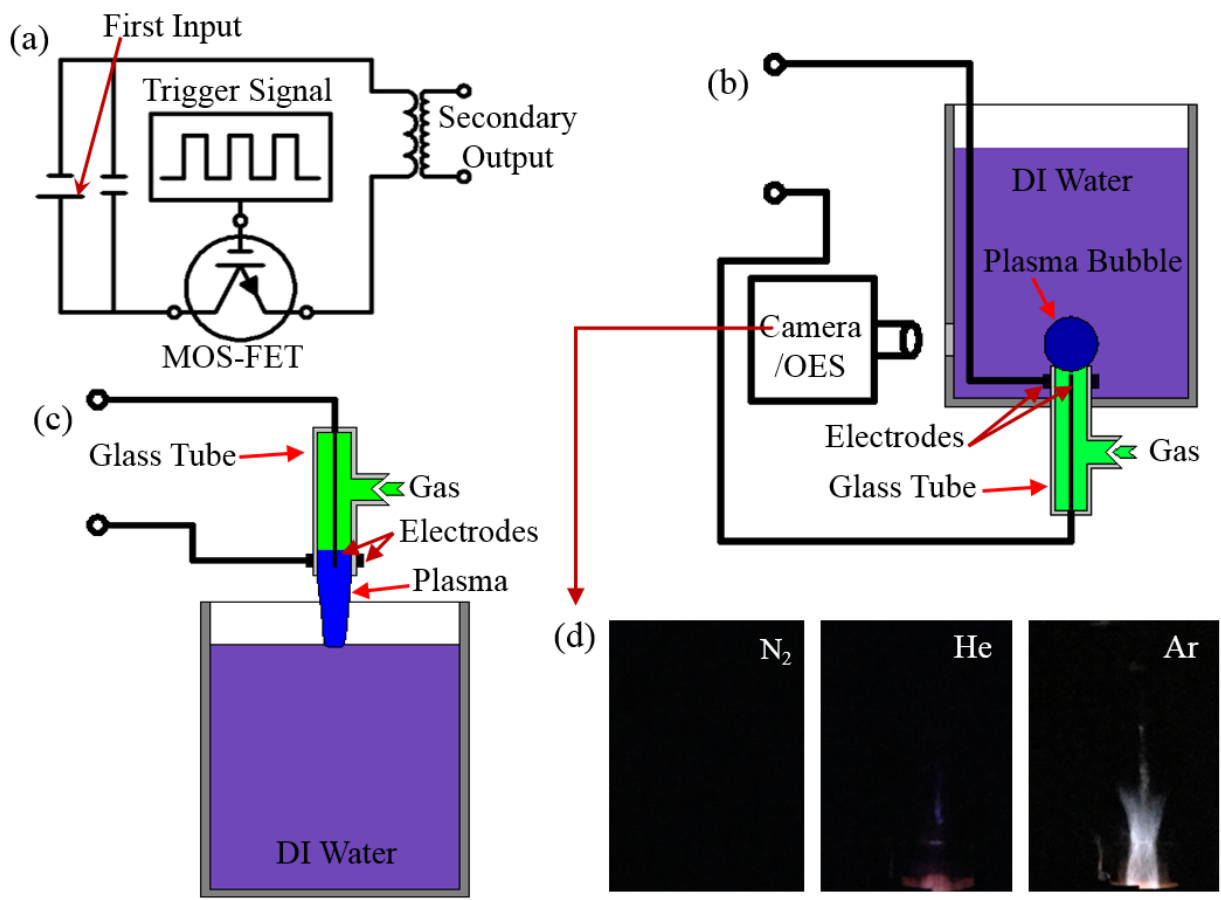

Figure 1. The cold atmospheric plasma device setup consists of a HV pulse generator (a) connected to a pin-to-plate electrode system submerged in DI water (b) and out of DI water (c). Rising bubbles (in DI water) or plasma jet (out of DI water) between the electrodes are produced by pumping a gas through a glass tube. Images of plasma generated in DI water using $N_2$, He, and Ar showing in (d).

*2.2 Optical Emission Spectroscopy*

Optical Emission Spectroscopy (OES) was used to assess various species (nitrogen [$N_2$], nitric oxide [–NO], nitrogen cation [$N^{+2}$], atomic oxygen [O], and hydroxyl radical [–OH]). Here, we show OES's results as qualitative analysis not quantitative research. In present work, UV-visible-NIR with a range of wavelength 200-850 nm was investigated on plasma discharged in water to detect various RNS and ROS. The spectrometer and the detection probe were purchased from Stellar Net Inc. In order to measure the radius of the plasma in DI water, a transparent glass plate was used to replace part of container. The optical probe was placed at a distance of 3.5 cm in front of plasma jet nozzle. Integration time of the collecting data was set to 100 ms.

*2.3 Measurement of RNS and ROS inside the solution*

For the comparative study among $N_2$, He and Ar, ROS and RNS levels in solution were analyzed after 30 min of plasma produced in 200 g DI water. RNS was measured by using the Griess Reagent System (Promega Corporation). All steps of detection were performed according to the manufacturer's instructions. The absorbance was measured at 540 nm by Synergy H1 Hybrid Multi-Mode Microplate Reader. Fluorimetric Hydrogen Peroxide Assay Kit was bought from Sigma-Aldrich for of $H_2O_2$ measurement. Detailed protocol can be found on the Sigma-Aldrich website. Briefly, we added 50 $\mu$l of standard curve samples, controls, and experimental samples to the 96-well flat-bottom black plates, and then added 50 $\mu$l of Master Mix to each of wells. We incubated the plates for 20 min at room temperature in dark condition and detected fluorescence by Synergy H1 Hybrid Multi-Mode Microplate Reader at Ex/Em: 540/590 nm.

*2.4 Breast cancer cell lines*

The human breast cancer cell line (MDA-MB-231) was provided by Dr. Zhang's lab at the George Washington University. Cells were cultured in Dulbecco's Modified Eagle Medium (DMEM, Life Technologies) supplemented with 10% (v/v) foetal bovine serum (Atlantic Biologicals) and 1% (v/v) penicillin and streptomycin (Life Technologies). Cultures were maintained at 37 °C in a humidified incubator containing 5% (v/v) $CO_2$. Cells were observed under a Nikon Eclipse TS100 inverted microscope.

*2.5 Cell viability assay*

MTT (3-(4, 5-dimethylthiazol-2-yl)-2,5-diphenyltetrazolium bromide) assay (Sigma-Aldrich) was used to monitored the cell viability. The cells were plated in 96-well flat-bottomed microplates at a density of 3000 cells in 100 ul of compete culture medium per well. Cells were the incubated for one day to ensure a proper cell adherence and stability. Confluence of each well was ensured to be at ~40%. Then we put 30 $\mu$l of DMEM, DI water, and plasma solutions to the wells to check effect of three mediums on the cells. After an additional incubation at 37 °C for 24 and 48 hours, the original culture medium was aspirated, and 100 $\mu$l of MTT solution per well (5 mg thiazolyl blue tetrazolium blue in 10 ml medium) was added into each well. After the subsequent 3 hours, the medium was replaced by 100 $\mu$l of MTT solvent ((0.4% (v/v)) HCl in anhydrous isopropanol) to dissolve formazan crystals. Then, the absorbance of solution was monitored by the Synergy H1 Hybrid Multi-Mode Microplate Reader at 570 nm.

## 3. Results and discussion

*3.1 Optical characteristics of plasma discharged in water using Ar, He, and $N_2$*

The reactive species generated by CAP discharged in DI water are shown in Fig. 2. The identification of emission lines and bands were performed according to reference[33]. Using nitrogen as a carries gas leads to formation of $N_2$ second-positive system ($C^3\Pi_u$-$B^3\Pi_g$) with its peaks at 316, 337, and 358 nm. There were very weak NO emission lines in the range of 250-300 nm[34]. When argon was used high-intensity OH peak at 309 nm[16] was observed as shown in Fig. 2c. Naturally argon and helium lines in the range of 600 to 800 nm were observed when argon and helium were used as shown in Fig. 2b and 2c. Overall higher emission intensity was found in the case of argon as a carries gas. OH peak intensity is almost the same as that of Ar in the case in argon plasma.

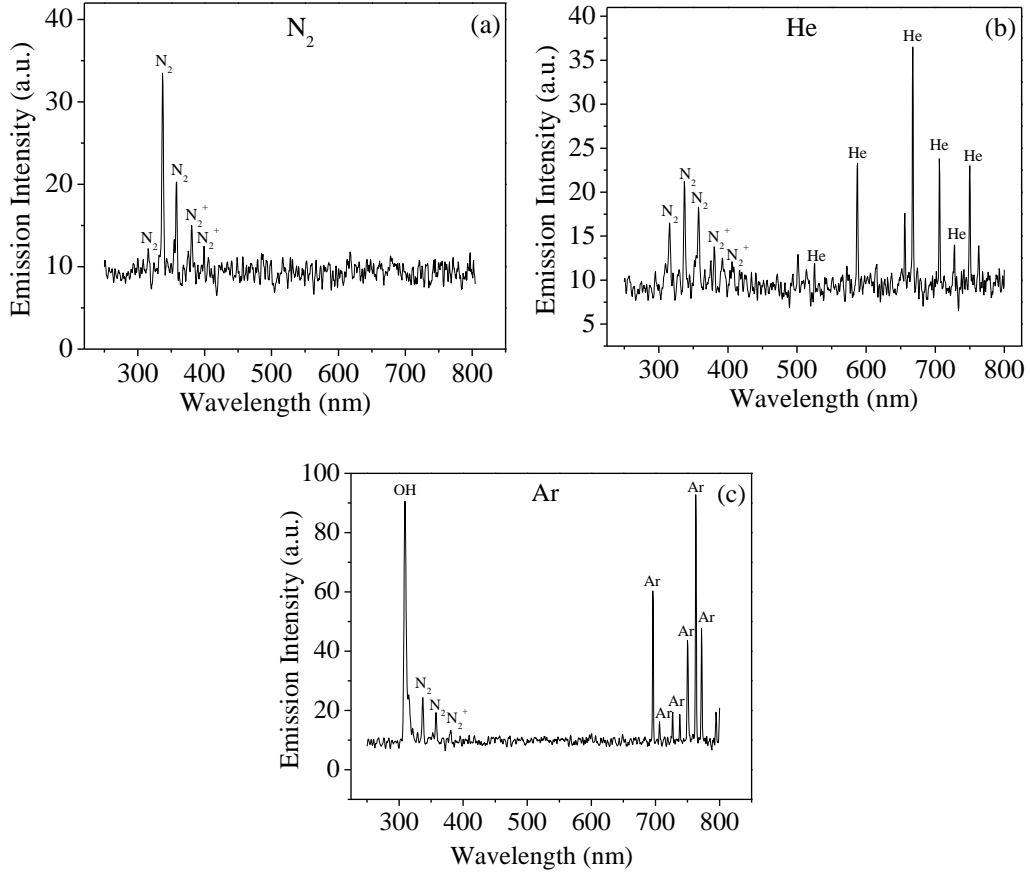

Figure 2. Emission spectrum of submerged CAP in DI water.

*3.2 Compare electron density of Ar, He, and N₂*

Due to limited access of submerged CAP, we measured the electron density of CAP setup shown in Fig 1c. Plasma density was measured using the Rayleigh microwave scattering (RMS), consisting of two microwave horns that were used for radiation and detection of microwave signal as shown in Fig. 3(a). Plasma density was obtained from the plasma conductivity by using the following expression: $\sigma = 2.82 \times 10^{-4} n_e v_m / (w^2 + v_m^2)$, $\Omega^{-1} cm^{-1}$, where $v_m$ is the frequency of the electron-neutral collisions, $n_e$ is the plasma density, and $w$ is the angular frequency[35]. The plasma conductivity, in turn, can be expressed in $U = A\sigma V$ with coefficient using $A = 263.8\ V\Omega/cm^2$, where $U$ is the output signal and $\varepsilon$ is the scatter constant[36,37]. The volume of the plasma column was determined from the Intensified Charged-Coupled Device (ICCD) images.

Temporal evolution of plasma density is presented in Fig. 3(b) to Fig. 3(d). One can see that electron density were about $2.3 \times 10^{11}$ cm$^{-3}$, $1.4 \times 10^{12}$ cm$^{-3}$, and $2.2 \times 10^{12}$ cm$^{-3}$ in the cases of N$_2$, He, and Ar, respectively.

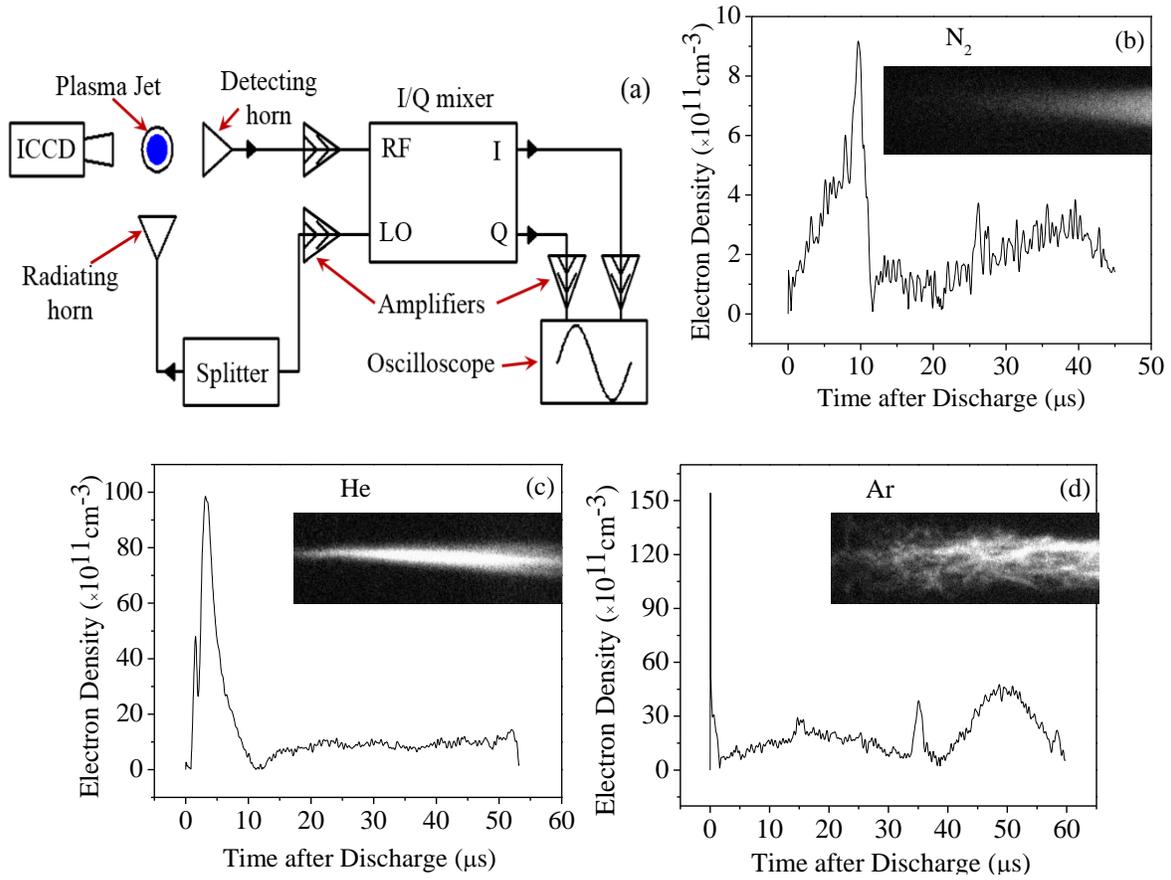

Figure 3. The schematics of RMS experimental setup (a). Temporal evolution of plasma density in N$_2$ (b), He (c) and Ar (d) atmospheric plasma jet. ICCD images of plasma using N$_2$, He, and Ar were insert into (b), (c), and (d), respectively.

*3.3 ROS and RNS in plasma solution*

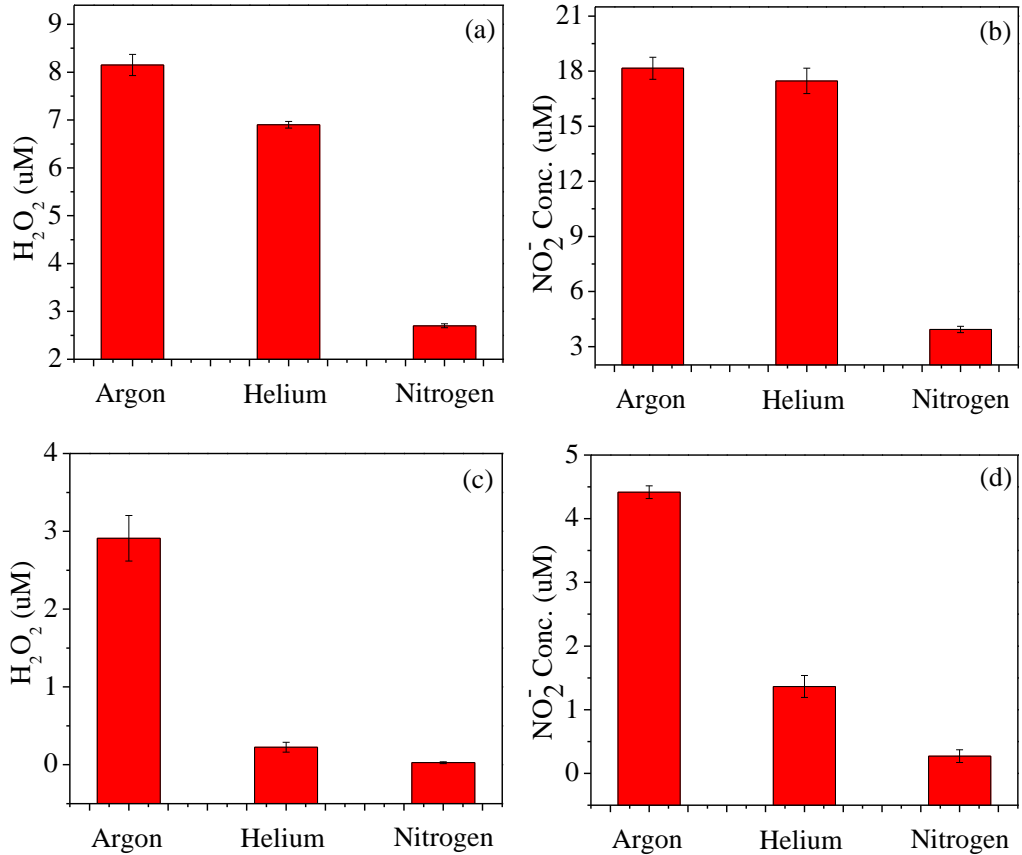

Figure 4. $H_2O_2$ and $NO_2^-$ concentration in plasma discharge in DI water. Figures 4a and 4b correspond to CAP submerged in DI water while Figs 4c and 4d correspond to CAP generated outside the water. $H_2O_2$ and $NO_2^-$ concentration are calculated by the concentration ratio of experimental group and control group. DI water volume is 200 ml.

Plasma discharged in water produces reactive species such as superoxide, the hydroxyl radical, singlet oxygen, and nitric oxide[38]. These relatively short-lived reactive species may be converted to relatively long-lived species such as hydrogen peroxide ($H_2O_2$), nitrite ($NO_x$), and other uncertain species[39]. $H_2O_2$ and $NO_x$ are known to induce cell proliferation as well as cell death. $H_2O_2$ is known to induce both apoptosis and necrosis, while $NO_x$ can induce cell death via DNA double-strands break[40,41]. On the other hand, Girard et al. indicated that $NO_2^-$ acts in synergy with $H_2O_2$ to enhance cell death in normal and tumor cell lines[42]. Therefore, we investigated effect of hydrogen peroxide ($H_2O_2$) and nitrite ($NO_x$) generated by plasma discharged in water on breast cancer cells.

In order to compare efficiency of RONS production in the case of submerged CAP device and CAP jet outside the DI water, we measured the concentration of $H_2O_2$ and $NO_2^-$. These results are shown in Fig. 4 with concentrations produced by submerged plasma shown in Figs. 4(a) and 4(b) and concentrations produced by CAP jet outside the DI water shown in Figs 4c and 4d. It can be noticed that $H_2O_2$ and $NO_2^-$ concentrations produced by CAP discharge in DI water are much higher than those produced by CAP jet. One can see that the $H_2O_2$ and $NO_2^-$ concentration are the highest in the case of argon as a carrier gas and lowest in the case of nitrogen. Recall that relatively large OH peak was observed in Ar gas plasma (as shown in Fig 2), this might be the reason for the higher $H_2O_2$ concentration. Despite a very short lifetime[43], hydroxyl (•OH) is one of the most important species in plasma produced solution[43]. In particular, •OH reaction leads to $H_2O_2$ formation[44]. Following mechanisms of hydroxyl formation in our cases can be suggested.

In the $N_2$ gas plasma, there are several additional:

$$e + H_2O \rightarrow H_2O^* + e \qquad (1)$$

$$H_2O^* \rightarrow •OH + H• \qquad (2)$$

$$N_2 + e \rightarrow 2N + e \qquad (3)$$

$$2N + H_2O \rightarrow N_2 + •OH + 2H• \qquad (4)$$

In the case of the He gas plasma, reactions (1) and (2) will remain the same but we have two additional reactions:

$$He + e \rightarrow He^* + e \qquad (5)$$

$$\text{He}^* + \text{H}_2\text{O} \rightarrow \text{He} + {}^\bullet\text{OH} + \text{H}^\bullet \quad (6)$$

In the case of Ar gas plasma, two additional reactions are possible:

$$\text{Ar} + \text{e} \rightarrow \text{Ar}^* + \text{e} \quad (5)$$

$$\text{Ar}^* + \text{H}_2\text{O} \rightarrow \text{Ar} + {}^\bullet\text{OH} + \text{H}^\bullet \quad (6)$$

Taking into account aforementioned chemistry it is plausible to explain dependence of $H_2O_2$ concentration on the carrier gas as shown in Fig. 4 (a).

Let's consider formation of $NO_2^-$ in CAP submerged in DI water. It is known that $NO_2^-$ is a primary breakdown product of NO in DI water[45]. In addition to the reaction mentioned above (i.e. Eq (3)) following are additional pathways leading to $NO_2^-$ production[38,46].

$$\text{N} + \text{O}_2 \rightarrow \text{NO} + \text{O} \quad (7)$$

$$4\text{NO} + \text{O}_2 + 2\text{H}_2\text{O} \rightarrow 4\text{NO}_2^- + 4\text{H}^+ \quad (8)$$

Owing to contact of DI water with air it is reasonable to assume that $O_2$ and perhaps $N_2$ is coming from air. Thus reactions (3), (7) and (8) can be used to explain the production of $NO_2^-$ in all three gases. As it is shown in Fig. 3, electron density is the highest in the case of argon plasma. Therefore, one can expect higher production of atomic nitrogen (N) leading to higher concentration of $NO_2^-$ in the case of argon plasma as shown in Fig. 4 (b).

*3.4 Effect of plasma solutions on cell viability*

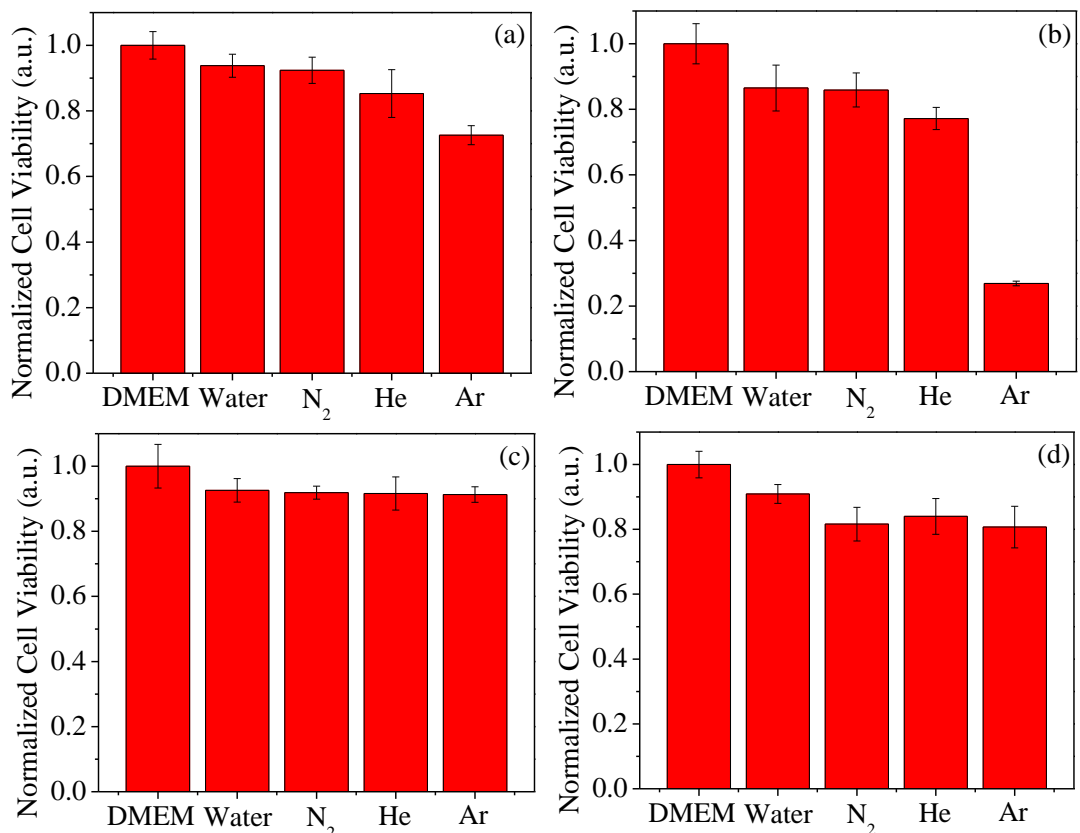

Figure 5. Effect of five solutions (DMEM, DI water, and plasma discharged in/out of DI water (using $N_2$, He, and Ar)) on cell viability. The human breast cancer cell line (MDA-MB-231) was treated by five solutions for 24 hours (plasma generated in DI water (a) and plasma generated out of DI water (c)) and 48 hours (plasma generated in DI water (b) and plasma generated out of DI water (d)) measured by MTT assay.

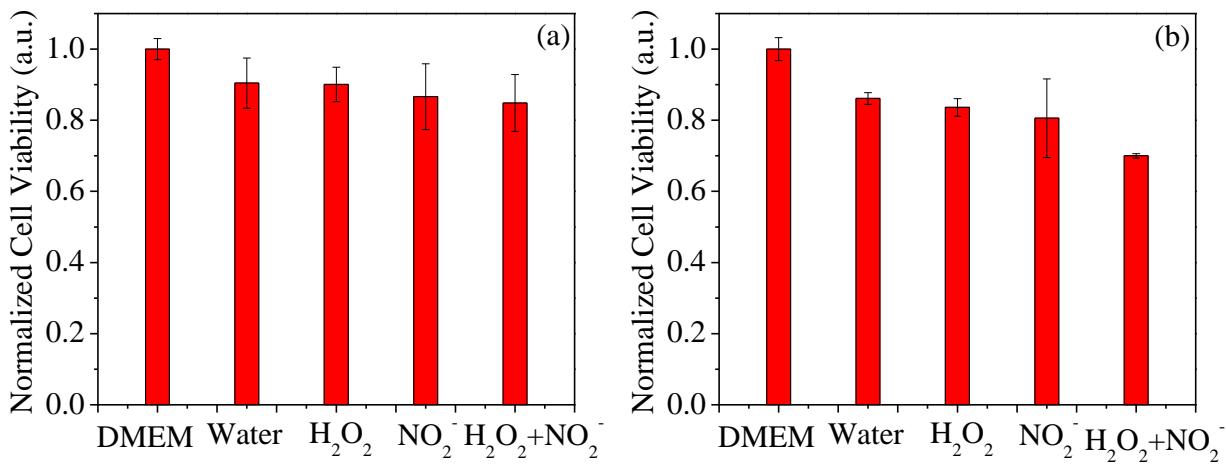

Figure 6. Effect of 8 uM $H_2O_2$, 18 uM $NO_2^-$, and 8 uM $H_2O_2$ + 18 uM $NO_2^-$ on cell viability using DMEM and DI water as control. The human breast cancer cell line (MDA-MB-231) was treated by five solutions for 24 hours (a) and 48 hours (b) measured by MTT assay.

Plasma discharged in DI water was applied to cancer cells. As control we used DMEM and untreated DI water. Fig. 5 shows the cell viability of the human breast cancer cell line exposed to DMEM, DI water, $N_2$ plasma solution, He plasma solution, and Ar plasma solution for 24 hours and 48 hours. The cell viability decreased by approximately 27.4% when treated with Ar plasma solution and 14.7% when treated with He plasma solution compared with DMEM solution when incubated for 24 hours (Fig. 5(a)). Only slight decrease of cell viability was observed in the case of DI water and $N_2$ plasma solution. Considering 48 hours' treatment, the viability decreased by approximately 73.1%, 22.8%, 14.1%, and 13.5% when cells treated with Ar plasma solution, He plasma solution, $N_2$ plasma solution, and DI water, respectively (Fig. 5(b)). Thus, the strongest effect can be observed in the case of Ar-feed plasma, while smallest effect was observed in the case of $N_2$-feed plasma. Comparing results of cell treatment with solution stimulated by plasma submerged in DI water (Fig. 5a, b) and plasma jet (Fig. 5c, d) lead to the conclusion that the strongest effect is in the case of Ar-feed plasma generated in DI water. Compared effects of $H_2O_2$ and $NO_2^-$ generated by plasma discharged in DI water on cancer cells with equal amounts of chemical, 24 and 48 hours' cell viability of 8 uM $H_2O_2$, 18 uM $NO_2^-$, and 8 uM $H_2O_2$ + 18 uM $NO_2^-$ were shown in Fig. 6. Synergistic effect of $H_2O_2$ and $NO_2^-$ on cancer cell is much better than individual. Comparing Fig. 5 and 6, one can see that Ar plasma solution exhibits a stronger cancer cell killing effect than equal amounts of $H_2O_2$ and $NO_2^-$ added.

In the study, it is shown that both ROS and RNS species are formed in plasma discharged in DI water. ROS are known to induce both apoptosis/necrosis, and RNS are capable of inducing the death of cells via DNA double-strands break (DDSB)/apoptosis[41,47]. It is interesting to point out that RONS measurements presented in Fig. 4 indicate that there is significant difference between concentration of $H_2O_2$ produced by Ar-feed plasma and He-feed plasma. While the concentration of $NO_2^-$ changes slightly in these two cases. This suggests that ROS and, in particular $H_2O_2$ play important role in interaction of plasma stimulated media with cells.

## 4. Conclusions

In summary, this study investigated effect of plasma discharged in DI water using Ar, He and $N_2$ gases as a carrier on breast cancer cells. Plasma generated in DI water using Ar has the strongest intensity while using $N_2$ is the weakest. The highest electron density of plasma is $2.2 \times 10^{12}$ cm$^{-3}$ also at Ar, which results into highest concentration of ROS and RNS. Compared with plasma generated in DI water, plasma generated out of DI water has lower concentration of ROS and RNS and weaker effect on cancer cells. Significant amount of ROS and RNS can be produced and argon was found to be the most efficient gas carrier to produces reactive solution to cause apoptosis in cancer cells. The results suggest the great potential of plasma generated in water for clinical cancer therapy.

**Acknowledgement.** This work was supported in part by a National Science Foundation, grant 1465061. Authors acknowledge Dr. A. Shashurin from College of Engineering at Purdue University for help in setting up RMS experiment and useful discussions. We thank Dr. Ka Bian and Dr. Ferid Murad from Department of Biochemistry and Molecular Medicine at The George Washington University for their support in the measurement of ROS and RNS experiments. We also thank Dr. Grace Zhang from Department of Mechanical and Aerospace Engineering at The Geroge Washington University for providing the breast cancer cells.